\documentclass{article} 

\usepackage{amsmath}
\usepackage{amssymb}
\usepackage{caption}
\usepackage{graphicx}
\usepackage{latexsym}

\begin{document}

\title{A minimal model for the role of the reaction rate on the initiation and self-sustenance of curved detonations}

\author{{\large Matei Ioan Petru R\u{a}dulescu$^{*}$, Aliou Sow}\\
        {\small \em Department of Mechanical Engineering, University of Ottawa, Canada}\\
        }

\date{}
\maketitle

\begin{abstract} 
A minimal model for curved detonations is studied, illustrating the role of the reaction rate on the detonation speed and its propagation limits.  The model is based on a simple extension of the minimal Fickett toy model for detonations based on the kinematic wave equation. The use of a simple depletion rate conditioned on the shock speed serves to illustrate its role in the quasi-steady structure of curved waves and their initiation from a strong blast wave. Calculations of strong initiation from a self-similar explosion illustrate the various asymptotic regimes of the transition to self-sustenance and their link to the steady wave structure.  We recover the asymptotic regimes of detonation formation suggested by He and Clavin and modelled in the context of Detonation Shock Dynamics by Stewart and collaborators. Following an analysis using the shock change equation, we identify a unique criterion that permits to infer the critical energy for initiation from the competition between energy release and geometric decay.   
\end{abstract}

%

\section{Introduction}
The propagation of curved detonation waves is a central problem of detonation theory.  The response of detonation waves to lateral boundary conditions when propagating in a weakly confined tube invariably leads to curved waves.  It has been established that an excess curvature leads to wave speed decrease, and possible extinction.  The problem has first been treated by Wood and Kirkwood \cite{wood1954}.  They recognized the presence of an internal sonic surface, along which the rate of exothermicity balances the rate of energy withdrawal due to divergence.  This simultaneous sonic condition and zero net energy evolution at the sonic surface is the so-called generalized Chapman-Jouguet condition. This confirmed that flow divergence acts similarly to heat losses, confirming the earlier findings of Zel'dovich and Kompaneets, who first demonstrated this generalized CJ condition \cite{zeldovich1960}.  The state of the art on the problem of curved detonations, as well as extensions to treat weakly non-steady effects are summarized by Bdzil and Stewart \cite{bdzil2007,bdzil2012}.  We provide experimental validation of the models for gaseous detonations in our previous studies where constant curvature and quasi-steady states were obtained in channels with exponentially enlarging cross-sectional areas \cite{radulescu2018dynamic, xiao2020dynamics, xiao2020role, zangene2023critical}.  

In the present communication, we wish to illustrate this problem with a minimal example.  Instead of using the Euler equations describing the reactive hydrodynamics of a reactive fluid, we instead use a reactive analogue problem first suggested by Fickett \cite{Fickett1979, Fickett1985}.  It consists of a kinematic wave equation coupled with energy release.  Using this toy model, we illustrate the role of the reaction rate liberating energy on the condition for self-sustenance, yielding the detonation speed - curvature relationship for quasi-steady propagation. Variations upon this toy model has been very useful in studying the stability and the non-linear dynamics of detonations \cite{radulescu2011nonlinear, kasimov2013model}. 

We also wish to study the role played by this relationship in the context of critically initiated detonations from a strong self-similar blast wave, a long standing problem in detonation theory \cite{lee_2008}. The relative simplicity of the model serves to illustrate and justify the model of He and Clavin \cite{he_clavin_1994}, subsequently refined by Kasimov and Stewart \cite{kasimov_stewart_2005}, Vidal \cite{vidal2009critical} and Clavin and Denet \cite{clavin_denet_2020} to account for transient effects.  The analysis of the dynamics are cast in the modern treatment of this problem using the concept of characteristics advocated by Kasimov and Stewart \cite{kasimov_stewart_2005} and the shock change equation closure for obtaining evolution equations for the front.  We extend the numerical study of Faria et al.\ on initiation of detonations in the context of the toy Fickett model \cite{faria2015qualitative}. 
 
\section{The Euler equations for quasi-1D or axi-symetric flow} 
Curved detonations can be described by the quasi-1D flow equations for conservation of mass, momentum and energy, which read
\begin{align}
\frac{\partial \rho}{\partial t}+\frac{\partial (\rho u)}{\partial x}&=-\rho u \kappa\\
\frac{\partial \rho u}{\partial t}+\frac{\partial (\rho u^2 +p)}{\partial x}&=-\rho u^2 \kappa\\
\frac{\partial \rho e_{tot}}{\partial t}+\frac{\partial (\rho u e_{tot} +pu)}{\partial x}&=-(\rho u e_{tot} +pu) \kappa
\end{align}
where $e_{tot}=e+\frac{1}{2}u^2$ is the total energy, which contains the reference energies of each component, contributing to the net heat release. We note that $\kappa$ is the rate of logarithmic area increase of the streamtube $\kappa=\frac{1}{A}\frac{\mathrm{d}A}{\mathrm{d}x}= \frac{\mathrm{d} \ln A}{\mathrm{d}x}$ and also the curvature of the front.  These equations are also those for cylindrical and spherical flows, with $\kappa=\frac{1}{A}\frac{\mathrm{d}A}{\mathrm{d}x} =j/x$, with $j$=1 or  2 for the cylindrical and spherical geometries respectively. 

\section{Fickett's model with lateral divergence} 
Fickett's model is an extension of the inviscid Burgers equation to model reactive compressible flow. A single toy model equation instead of three conservation laws for the Euler equations describes the compressible hydrodynamics. In the absence of lateral divergence, the model is
\begin{equation}
\frac{\partial \rho }{\partial t}+\frac{\partial p}{\partial x}=0 \label{eq:fickettideal}
\end{equation}
where $p$ corresponds to the flux of the quantity for which $\rho$ is its density.  Fickett models the flux by:
\begin{equation}
p = \frac{1}{2}\left( \rho^2 + \lambda Q \right) \label{eq:flux}
\end{equation} 
where $Q$ is the energy release and $\lambda$ is a progress variable 0 in the reactants and 1 in the products.  In doing so, the model equation accounts for the modification of the flux by the energy release. In the absence of the reactive term, it is the inviscid Burgers equation. Fickett ascribes the meaning of $p$ to something akin to pressure.  This reflects the notion that the flux is related to pressure in the real system.  The mathematical structure of this toy model retains the basic physics of the reactive Euler equations, but it is much simpler to analyze its emerging dynamics. It thus serves very well its main purpose of a toy model.

This simple partial differential equation has to be supplemented by another dictating the rate of reaction.  Fickett uses 
\begin{equation}
\frac{\partial \lambda }{\partial t}=r \label{eq:ratepde}
\end{equation}
where $r$ is the reaction rate. 

Our simple extension to treat curved detonations is to consider the wave propagation in a channel of enlarging cross section $A(x)$.  For such quasi-one dimensional flow, a simple extension of Fickett's model \eqref{eq:ficketoriginal} is:
\begin{equation}
\frac{\partial \rho }{\partial t}+\frac{\partial p}{\partial x}=-\frac{1}{2}\rho^2 \kappa \label{eq:fickettloss}
\end{equation}

This retains the structure of the reactive Euler equations, where a non-linear term, usually the flux of the conserved quantity (except for momentum) appears on the right hand side, multiplying the curvature.  

In the following problem, we take the rate as 
\begin{equation}
r=D^n\left(1-\lambda\right) \label{eq:rate}
\end{equation}
where $D$ is the detonation speed.  This choice is motivated by the physical picture where most detonations have a thin induction zone followed by a hydrodynamic structure, which, one average, is two orders of magnitude longer.  By eliminating the induction zone, we also remove the pulsating instability that would otherwise obscure the slow dynamics we wish to unravel. We allow the sensitivity of the reaction rate on the shock strength to be described by the exponent $n$.  The same choice was made by Fickett in his discussion of \textit{steady} detonations with lateral divergence, but in which he chose a more artificial model equation:
\begin{equation}
\frac{\partial \rho }{\partial t}+\frac{\partial p}{\partial x}=- \kappa  \label{eq:ficketoriginal}
\end{equation}
instead of our starting point \eqref{eq:fickettloss}. 

%

\section{Characteristic form}
It is straightforward to cast our problem given by \eqref{eq:ratepde}  and \eqref{eq:fickettloss} in characteristic form:
\begin{align}
\frac{\partial p }{\partial t}+\rho \frac{\partial p}{\partial x}&=\frac{1}{2}rQ-\frac{1}{2}\rho^3\kappa \label{eq:char1}\\
\frac{\partial \lambda }{\partial t}&=r \label{eq:char2}
\end{align}
where the characteristics speeds are respectively $\rho$ and 0.  We label the characteristics in the first equation given by the paths satisfying $dx/dt=\rho$ as C+ characteristics, by analogy to the physical system. 
\section{The generalized CJ condition based on characteristics}
\begin{figure}
	\center
	\includegraphics[width=0.8\columnwidth]{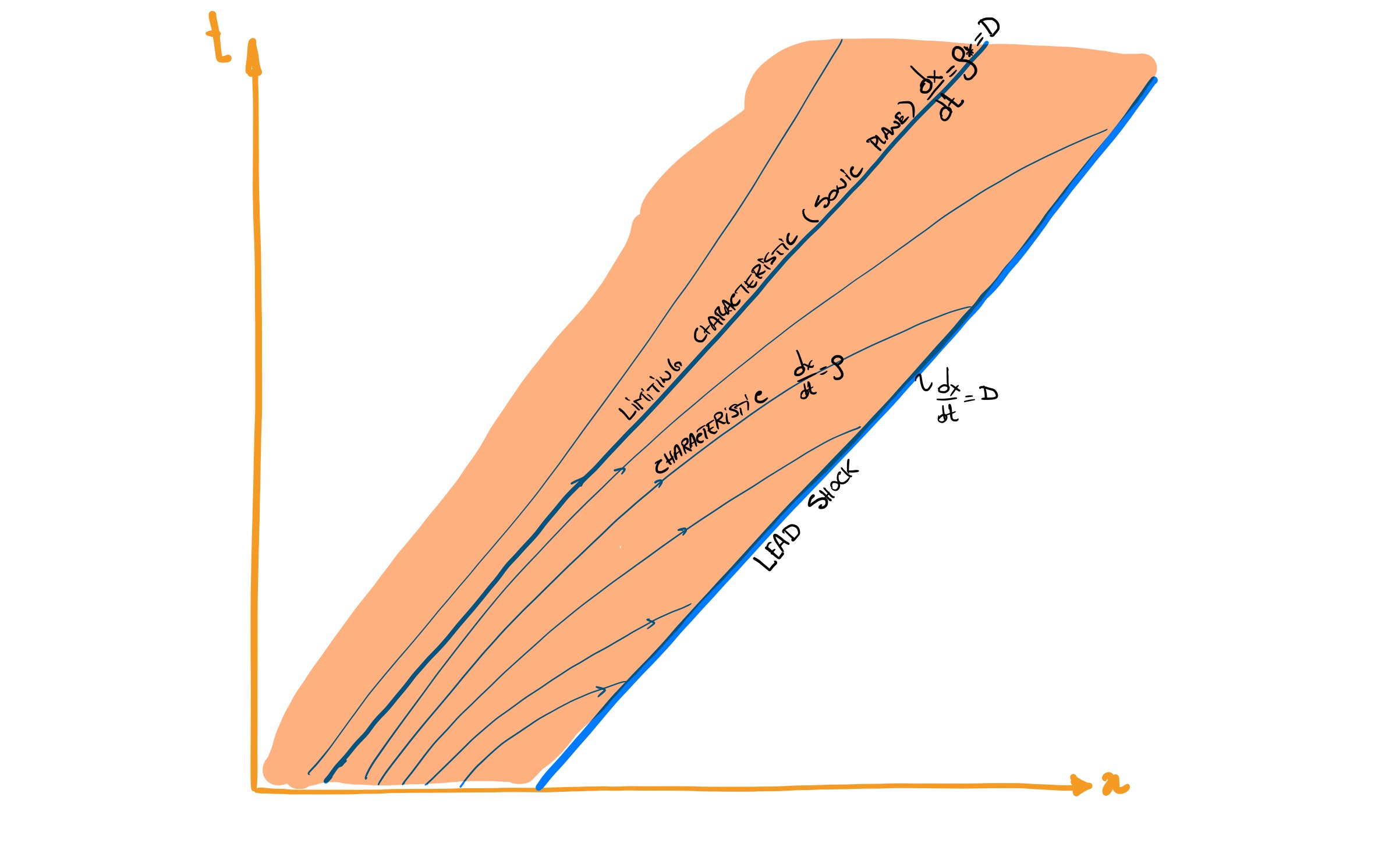}
	\caption{The structure of a steady travelling detonation wave and the network of forward facing characteristics sustaining the lead shock.}
	\label{fig:characteristics}
\end{figure}

\addvspace{10pt}
 The role of these characteristics for the analogue problem has been discussed by Radulescu and co-workers \cite{radulescu2011nonlinear, tang2013dynamics, lau2019multiplicity}, in analogy to the physical system\cite{kasimov_stewart_2005}.  For a steady traveling wave with speed $D$, characteristics from the back feed into the lead shock, providing an amplification of the pressure $p$ dictated by the rate of energy release in the first term on the right hand side of \eqref{eq:char1} and a reduction of the pressure by the second term on the right hand side of \eqref{eq:char1}. The steady structure can thus be simply viewed as the coherent amplification of forward facing waves by the energy release set up by the shock and modulated by the geometrical terms.  Figure \ref{fig:characteristics} illustrates this structure.  

The limiting characteristic is the characteristic where the flow is sonic in the frame of the lead shock.  It corresponds to the characteristic path parallel to the lead shock, such that 
\begin{equation}
\rho_*=D \label{eq:CJ1}\\
\end{equation}
For this characteristic to remain at constant speed, it requires that $p$ also remains constant along its path.  Since $\lambda$ is also a constant along this path by the constraint provided by the reaction rate in the steady travelling wave by virtue of \eqref{eq:char2}, \eqref{eq:char1} requires its right hand side forcing to vanish: 
\begin{equation}
\frac{1}{2}r_*Q-\frac{1}{2}\rho_*^3 \kappa =0 \label{eq:CJ2}
\end{equation}
The sonic condition \eqref{eq:CJ1} and vanishing of the net rate controlling flux amplification \eqref{eq:CJ2} is the generalized Chapman-Jouguet condition.

Using the expressions for the rate \eqref{eq:rate} and the model flux \eqref{eq:flux}, the   balance given by \eqref{eq:CJ2} gives the progress variable at the sonic surface, i.e.,
\begin{equation}
\lambda_*=1-\frac{D^{3-n} \kappa}{Q} \label{eq:lambdastar}
\end{equation}
The generalized CJ condition thus limits the amount of energy release felt by the shock to the fraction $\lambda^*$, which is less than unity.  This is given by the balance of the reaction rate and rate of loss at the sonic surface. 

\section{The inner structure as a BVP}
The detonation wave speed is determined from requiring that the internal steady structure satisfies the post shock state and sonic state simultaneously.  The sonic state is known from above, i.e, $(\rho=\rho_*=D, \lambda=\lambda_*)$.  The post-shock state without energy evolution is obtained from the weak solution of the inviscid Burgers equation.  
\begin{equation}
D=\frac{[p]}{[\rho]}
\end{equation}
where the square brackets denote differences across the shock wave.  Since we take the strong shock approximation with $p=\rho=0$ in the upstream state, the wave speed relation becomes:
\begin{equation}
D=\frac{\frac{1}{2} \rho^2}{\rho}=\frac{1}{2} \rho
\end{equation}
For a given detonation speed $D$, the post-shock density is thus $\rho=2D$.  At the shock, $\lambda=0$.  

The steady structure of the detonation is obtained by reverting to a wave-fixed reference frame and requiring the structure to be steady.  Let the new coordinate system $(\zeta, t')$ be related to the $(x,t)$ by:
\begin{align}
\zeta=x-\int^t_0D(\alpha)\mathrm{d}\alpha\\
t'=t
\end{align}
where $D$ is the shock speed. 
The derivatives appearing in \eqref{eq:ratepde}  and \eqref{eq:fickettloss} transform accordingly as:
\begin{align}
\partial_{x,t}=\partial_{\zeta,t'} \partial_{x,t}\zeta+\partial_{t',\zeta} \partial_{x,t}t'=\partial_{\zeta,t'}\\
\partial_{t,x}=\partial_{\zeta,t'} \partial_{t,x}\zeta+\partial_{t',\zeta} \partial_{t,x}t'=-D\partial_{\zeta,t'}+\partial_{t',\zeta}
\end{align}
Using this coordinate transformation, \eqref{eq:ratepde}  and \eqref{eq:fickettloss} become:
\begin{align}
\frac{\partial \rho }{\partial t'} & - D \frac{\partial \rho }{\partial \zeta} +\frac{\partial p}{\partial \zeta}=- \frac{1}{2}\rho^2 \kappa\\
\frac{\partial \lambda }{\partial t'} & - D \frac{\partial \lambda }{\partial \zeta} = r
\end{align}
The steady travelling wave solution corresponds to taking $\frac{\partial }{\partial t'}=0$.  We are left with two ordinary differential equations:
\begin{align}
 - D \frac{\mathrm{d} \rho }{\mathrm{d} \zeta} +\frac{\mathrm{d} p}{\mathrm{d} \zeta}=- \frac{1}{2}\rho^2 \kappa\\
- D \frac{\mathrm{d} \lambda }{\mathrm{d} \zeta} = r
\end{align}
These can be re-written as:
\begin{align}
\frac{\mathrm{d} \rho }{\mathrm{d} \zeta} =\frac{\frac{1}{2}Q\frac{r}{D} -\frac{1}{2}\rho^2 \kappa}{\rho-D} \label{eq:innerBVP1}\\
\frac{\mathrm{d} \lambda }{\mathrm{d} \zeta} = -\frac{r}{D} \label{eq:innerBVP2}
\end{align}
The problem we thus wish to solve is a boundary value problem given by \eqref{eq:innerBVP1} and \eqref{eq:innerBVP2}, satisfying the boundary conditions at the shock and at the sonic surface.  Cast in this form, we recognize again the requirement for the generalized CJ condition, requiring the simultaneous sonic flow (denominator vanishing) and balance of exothermic and endothermic effects (numerator vanishing) at the sonic surface.    While mathematically it is equivalent with what we already established above using the limiting characteristic criterion, we prefer the latter, as being more physical.  

The rate equation \eqref{eq:innerBVP2} readily integrates to 
\begin{align}
\lambda=1-\exp\left( D^{n-1}\zeta \right)
\end{align}
from which we can deduce the reaction zone length using \eqref{eq:lambdastar}.

\begin{figure}
	\center
	\includegraphics[width=0.8\columnwidth]{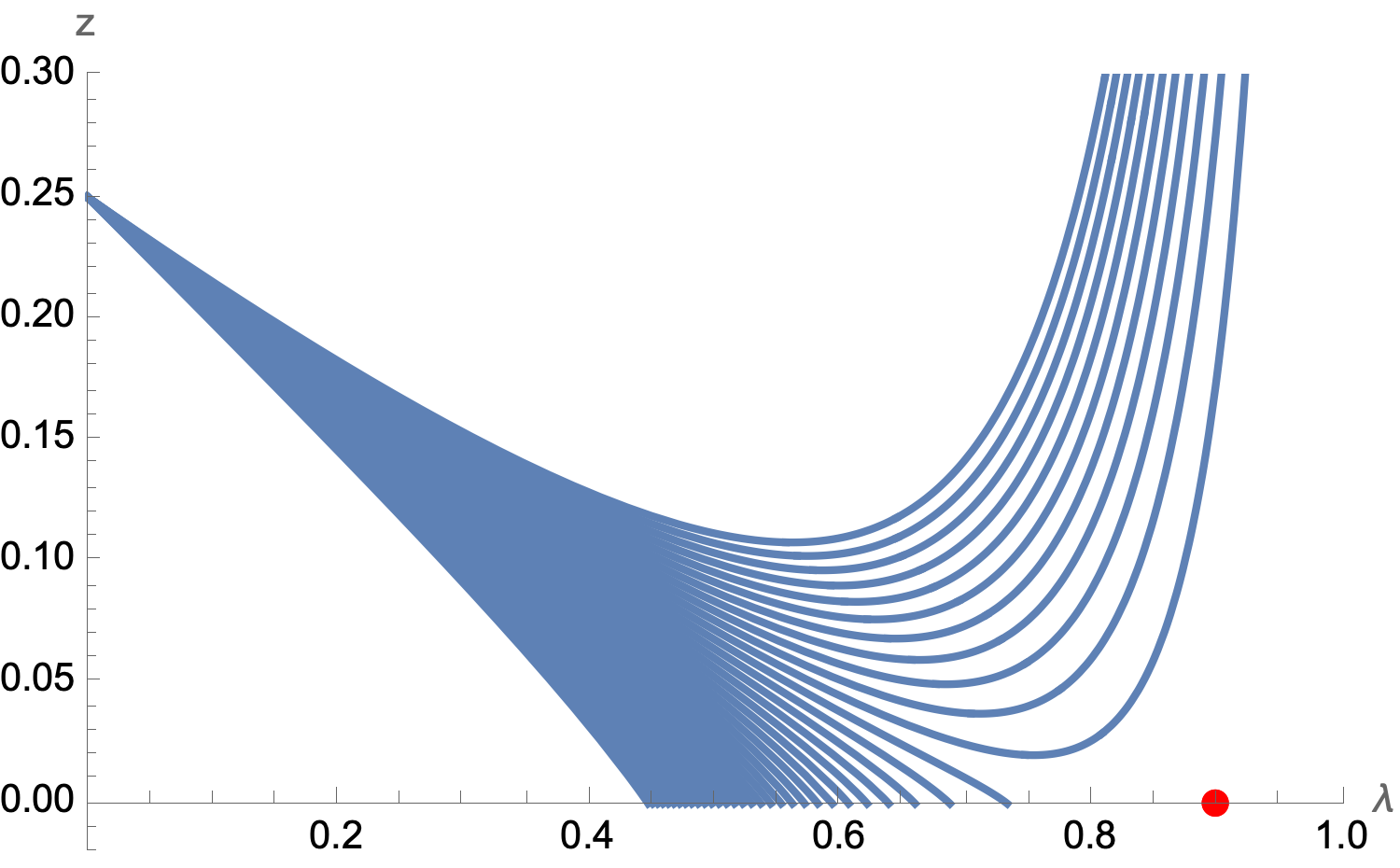}
	\caption{Phase space trajectory of the solution for fixed $D$ and varying $\kappa$; red circle denotes the sonic point found numerically.}
	\label{fig:saddle}
\end{figure}

The remaining ODE \eqref{eq:innerBVP1} can be manipulated by multiplying by $\rho-D$, dividing by \eqref{eq:innerBVP2}, and introducing the change of variables 
\begin{equation}
z=(\rho-D)^2 \label{eq:z}
\end{equation}
such that $(\rho-D)\mathrm{d}\rho = \frac{1}{2}\mathrm{d}z$.  It yields:
\begin{align}
\frac{\mathrm{d} z}{\mathrm{d} \lambda}=-Q +\frac{(D+\sqrt{z})^2\kappa}{D^{n-1}(1-\lambda)} \label{eq:BVP3}
\end{align}
At the shock, $z=D^2$, while $z=0$ at the sonic surface.  Using \eqref{eq:lambdastar}, it can be shown that the sonic locus corresponds to $\frac{\mathrm{d} z}{\mathrm{d} \lambda}=0$.  The sonic point is a degenerate saddle point, as shown in Fig.\ \ref{fig:saddle}.  For given shock speed, trajectories for less than critical values $\kappa$ approach the $z=0$ axis on the left of the sonic point with finite slope.  For values of $\kappa$ sufficiently large, trajectories reach zero gradient prior to reaching the sonic point.  The critical value of $\kappa$ is obtained for trajectories reaching the $z=0$ axis with zero slope.  This marks the eigenvalue $\kappa(D)$. 

A closed form solution was not possible since the boundary value problem given by \eqref{eq:BVP3} cannot be integrated in closed form.  This is usually the case in all problems of curved detonations.  Instead, the BVP can be solved numerically very simply by a shooting method, using the bevavior near the sonic point as an indicator.  Alternatively, linearization about the saddle point can determine the initial direction for a backwards shooting method towards the shock.  For the present case, the change of variables from $\rho$ to $(\rho-D)^2$  eliminated the blow-up at the sonic surface, rendering forward or backward shooting methods equivalent.  The change of variables \eqref{eq:z} removing the singularity at the sonic point was first identified by Fickett \cite{Fickett1985} while studying a simpler analog problem. It was exploited further by Faria and his collaborators for a model similar to ours \cite{faria2015qualitative} and generalized to the Euler equations as well, simplifying the numerical determination of the inner structure of detonations with losses \cite{semenko2016}.  

\begin{figure}
	\center
	\includegraphics[width=1\columnwidth]{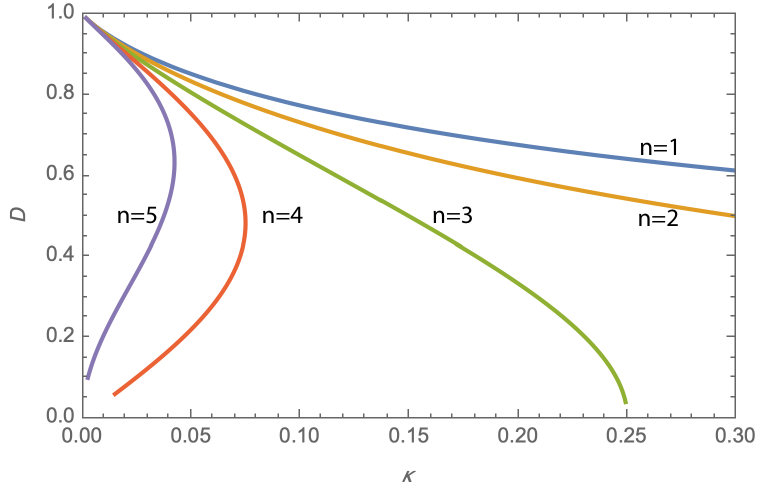}
	\caption{Steady detonation speed dependence on curvature for different values of $n$.}
	\label{fig:Dkappa}
\end{figure}

Figure \ref{fig:Dkappa} shows the numerically obtained $D(\kappa)$ curve for different values of $n$.  For $n$ larger than 3, the curves display the characteristic turning point at the maximum permissible curvature.  This is in very good qualitative agreement with the physical system. 

\section{Fickett's simplest model for losses and divergence effects}
It is worthwhile comparing our results with those of Fickett \cite{Fickett1985}, who investigated the structure of detonations given by \eqref{eq:ficketoriginal}:
\begin{equation*}
\frac{\partial \rho }{\partial t}+\frac{\partial p}{\partial x}=- \kappa  \end{equation*}

Proceeding as we have illustrated above, the characteristic form of the problem is given by 
\begin{align}
\frac{\partial f }{\partial t}+\rho \frac{\partial p}{\partial x}&=\frac{1}{2}rQ-\rho \kappa \label{eq:charfickettoriginal}
\end{align}
instead of \eqref{eq:char1}.  From this equation we deduce the generalized CJ condition $\rho_*=D$ and 
\begin{equation}
\lambda^*=1-2 \kappa D^{1-n} \label{eq:lambdastaroriginal}
\end{equation}
instead of \eqref{eq:lambdastar}.  The eigenvalue $\kappa$ is obtained again by requiring the integral curve connecting the post shock state and the sonic point.  With the simpler loss term assumed here, we obtain
\begin{align}
\frac{\mathrm{d} z}{\mathrm{d} \lambda}=-Q +\frac{2 \kappa}{D^{n-1}(1-\lambda)} \label{eq:BVP4}
\end{align} 
instead of \eqref{eq:BVP3}. The integral curve is readily integrated in closed form in this case by separation of variables:
\begin{align}
\int_{D^2}^0 \mathrm{d} z = \int_{0}^{\lambda^*} \left( -Q +\frac{2 \kappa}{D^{n-1}(1-\lambda)}\right) \mathrm{d} \lambda \label{eq:BVP5}
\end{align} 
yielding:
\begin{equation}
D^2=\lambda^* Q+\frac{2 \kappa}{D^{n-1}}\ln\left(1-\lambda^*  \right)
\end{equation}
This shows clearly the two contributions to the velocity deficit in the presence of curvature.  The first term accounts for the incomplete energy release, while the second accounts for the effect of curvature in modifying the location of the sonic surface in relation to the shock. 

With $\lambda^*$ given by \eqref{eq:lambdastaroriginal}, the final $D(\kappa)$ dependence is found to satisfy the transcendental relation
\begin{equation}
D^2=\left( 1-2 \kappa D^{1-n} \right) Q+2 \kappa D^{1-n}\ln\left(2 \kappa D^{1-n}  \right)
\end{equation}

\section{Direct initiation of spherical waves} 
We now turn to illustrating the utility of concepts used to obtain the steady detonation curvature response in non -steady situations.  We focus on determining the critical conditions for detonation initiation from the decay of a strong self-similar blast wave.  This is the analogous problem to the well-known direct initiation problem of detonation, where an initially strong self-similar Taylor-Sedov blast wave decays in a reactive gas \cite{lee_2008}.  For sufficiently large energy deposition, the blast decays but rapidly re-amplifies to form a detonation.  Below a critical value of energy deposition, the blast does not re-amplify and a detonation is not initiated. 
\begin{figure}
	\center
	\includegraphics[width=1\columnwidth]{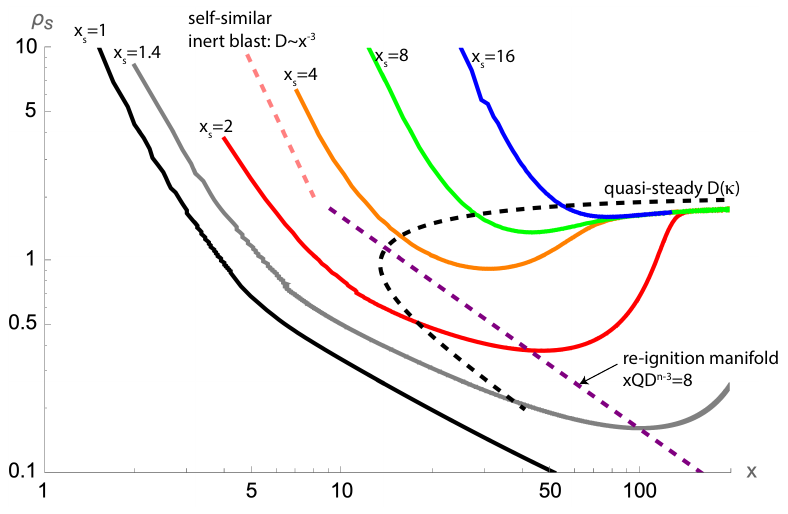}
	\caption{Shock density evolution for blasts of different strengths for $n=4$ and $Q=1$.}
	\label{fig:logdecays}
\end{figure}
\begin{figure}
	\center
	\includegraphics[width=\columnwidth]{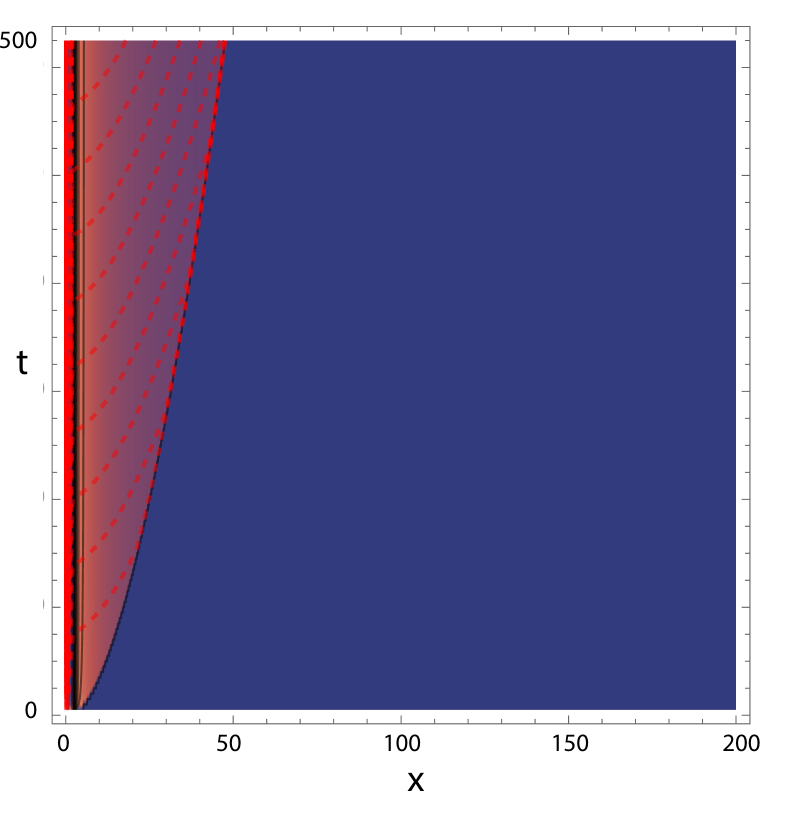}
	\caption{Space time diagrams for the evolution of the density field for $x_s=1$ ; characteristic trajectories are in red while $\lambda$ contours in black.}
	\label{fig:xts1}
\end{figure} 

\begin{figure}
	\center
	\includegraphics[width=\columnwidth]{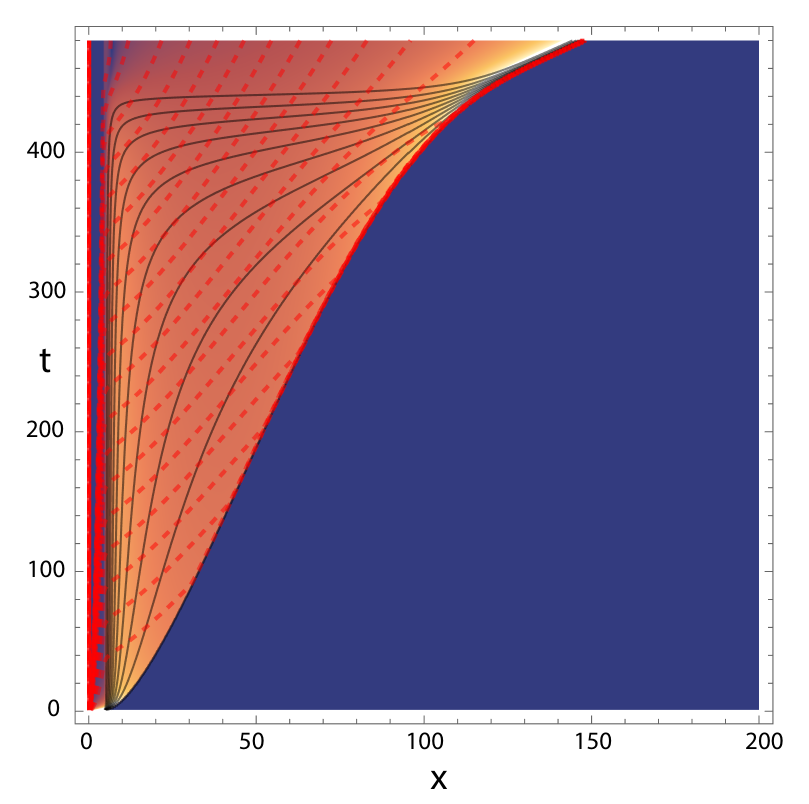}
	\caption{Space time diagrams for the evolution of the density field for $x_s=2$; characteristic trajectories are in red while $\lambda$ contours in black.}
	\label{fig:xts2}
\end{figure} 

\begin{figure}
	\center
	\includegraphics[width=\columnwidth]{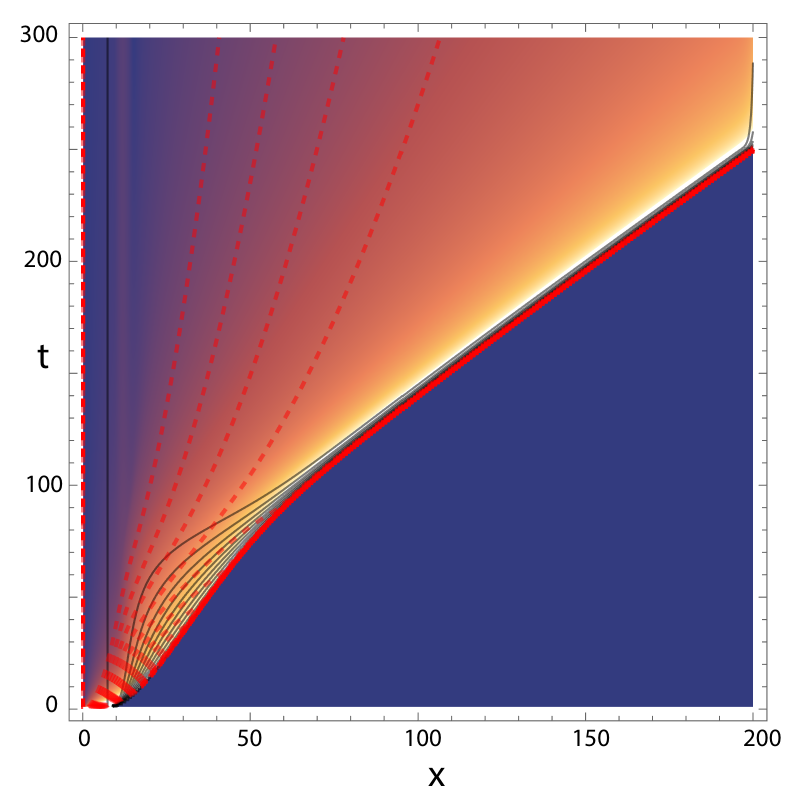}
	\caption{Space time diagrams for the evolution of the density field for $x_s=4$; characteristic trajectories are in red while $\lambda$ contours in black.}
	\label{fig:xts4}
\end{figure} 

\begin{figure}
	\center
	\includegraphics[width=1\columnwidth]{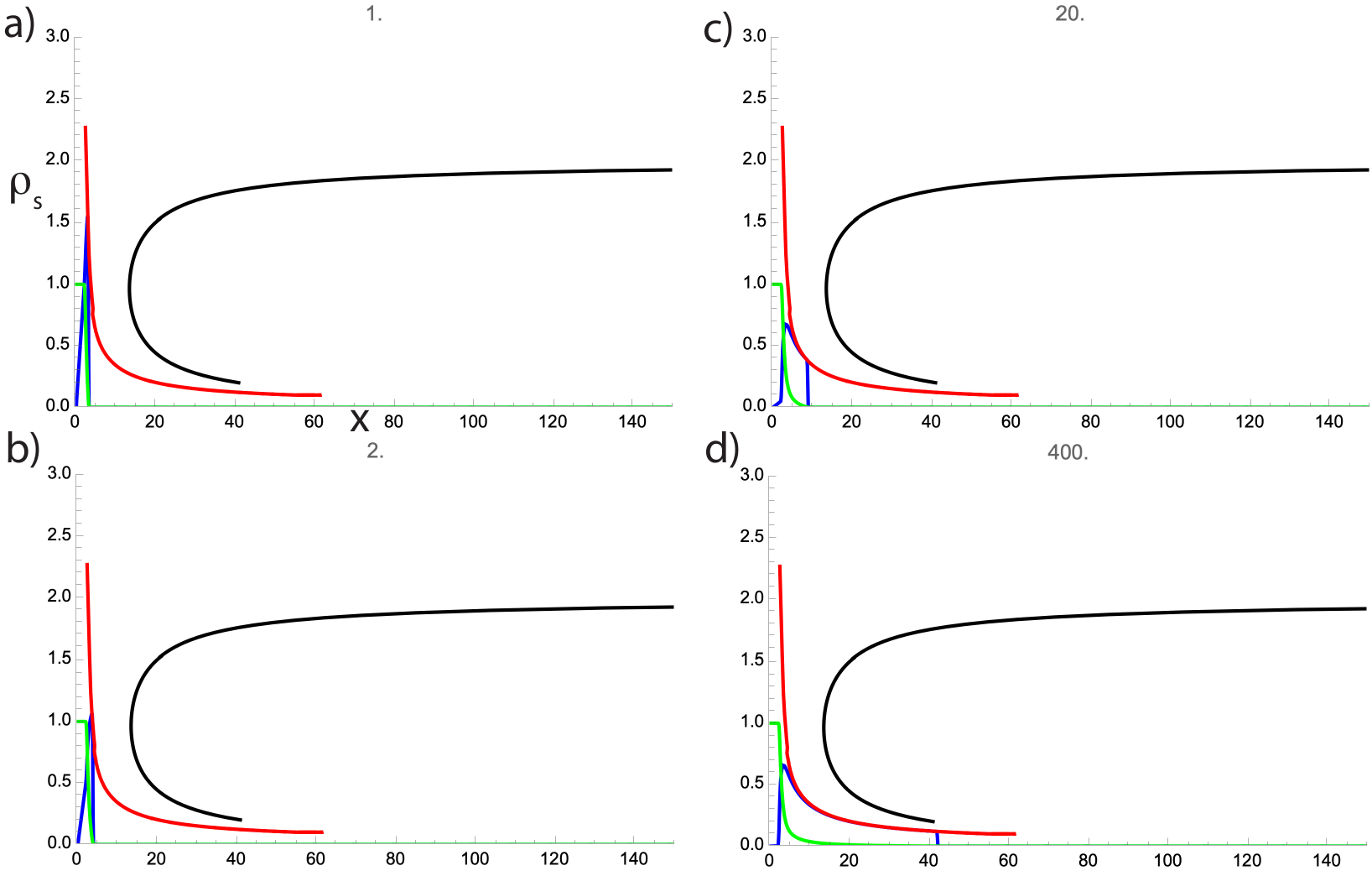}
	\caption{The fields $\rho(x)$ (blue), $\lambda(x)$ (green), $\rho_s(x)$ and the steady response (black) for $x_s = 1$ at a) $t=1$, b) $t=2$, c) $t=20$ and d) $t=400$; see also animations as Supplemental Material.}
	\label{fig:x1}
\end{figure} 

\begin{figure}
	\center
	\includegraphics[width=1\columnwidth]{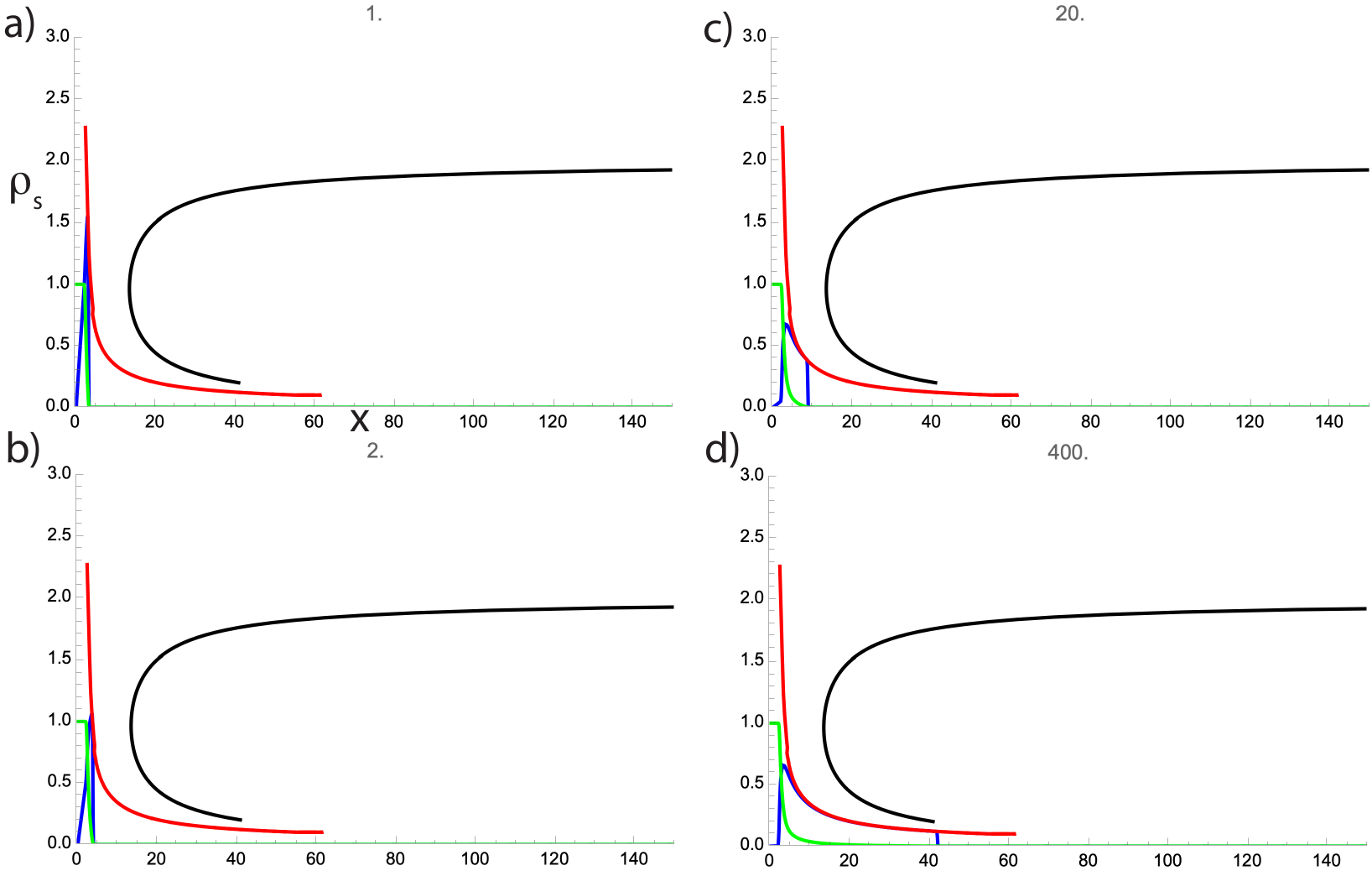}
	\caption{The fields $\rho(x)$ (blue), $\lambda(x)$ (green), $\rho_s(x)$ and the steady response (black) for $x_s = 2$ at a) $t=1$, b) $t=4$, c) $t=30$ and d) $t=400$; see also animations as Supplemental Material.}
	\label{fig:x2}
\end{figure} 
The problem we wish to solve is the direct initiation of spherical waves ($j=2$), and the curvature is $\kappa=j/x=2/x$ in \eqref{eq:fickettloss}, re-written here: 
\begin{equation}
\frac{\partial \rho }{\partial t}+\frac{\partial \left(\frac{1}{2}\rho^2+ \frac{1}{2}\lambda Q \right)}{\partial x}=-\frac{j}{x} \left(\frac{1}{2}\rho^2\right)  \label{eq:ficketspherical}
\end{equation}
At early times, a point source deposition of mass $m_0$ imposes a self-similar shock decay analogous to the Taylor-Sedov blast wave resulting from a point source "energy" deposition.  In the burgers case, the shock sphere conserves the mass integral \cite{radulescu2023self}.  Taking the limit $\rho^2 \gg Q$, the problem is an inert one at early times.  The inert point blast problem for the Fickett equation admits a very simple self-similar solution \cite{radulescu2023self}given by 
\begin{align}
\rho=2D\frac{x}{x_s}
\end{align}
where $x_s(t)$ is the position of the shock.  For the spherical problem, the shock speed decay is given in terms of the total mass $m_0$ deposited at the center by 
 \begin{align}
 D=\frac{m_0x^{-3}}{2\pi}
 \end{align}
The self-similar solution serves as our initial condition.  

To explore the complex dynamics embedded in this seemingly simple problem, we have first obtained numerical solutions to this reactive problem by an explicit simple finite volume scheme discretizing the flux by the Roe method \cite{radulescu2011nonlinear}. The geometric and reaction terms were integrated by operator splitting.  As initial condition, we impose the self-similar inert solution with a shock amplitude of $\rho_{s,0}=40$ and vary the location of the shock $x_{s,0}$ to bracket the critical initiation conditions. We focus on the case $n=4$ and take without loss of generality $Q=1$.

Figure \ref{fig:logdecays} shows the evolution of the post shock density $\rho_s$ for different blast intensities. Also shown is the steady state response obtained above, reported here in terms of the radius of curvature.  Recall that for a strong Burgers shock with $\rho=0$ upstream, the post shock density is $\rho_s=2D$.  All blasts decay with the expected self-similar decay of $x^{-3}$ until $\rho^2 \sim Q$, when the energy release contribution slows down the decay rate of the blast.  For reference, the Chapman Jouguet solution corresponds to $\rho_s=2D_{CJ}=2\sqrt{Q}$.

Depending on the magnitude of the blast, we recover two types of solutions.  The trajectories for blasts originating at a source radius of 4 or greater (for the fixed strength imposed) decay below the steady response and quickly re-amplify to form a self-sustained detonation. These asymptote towards the quasi-steady self-sustained response. For blasts originating at a source radius below 2, the decay is very long, with re-amplification occurring over very long times, rapidly tending to very large distances as the blast is reduced in strength.  For example, the $x_s=1$ trajectory does not re-amplify in our domain that is 200 long. 

The critical behaviour thus appears to correlate very well with the passage of the integral curves close to the steady state turning point.  Our model is thus in perfect accord with the results of numerical simulations in the reactive Euler equations as well \cite{he1996theoretical, watt_sharpe_2005}, as well as Faria's study of the analog equations with an induction model \cite{faria2015qualitative}, in spite of pulsating instabilities masking the dynamics.  
\begin{figure}
	\center
	\includegraphics[width=1\columnwidth]{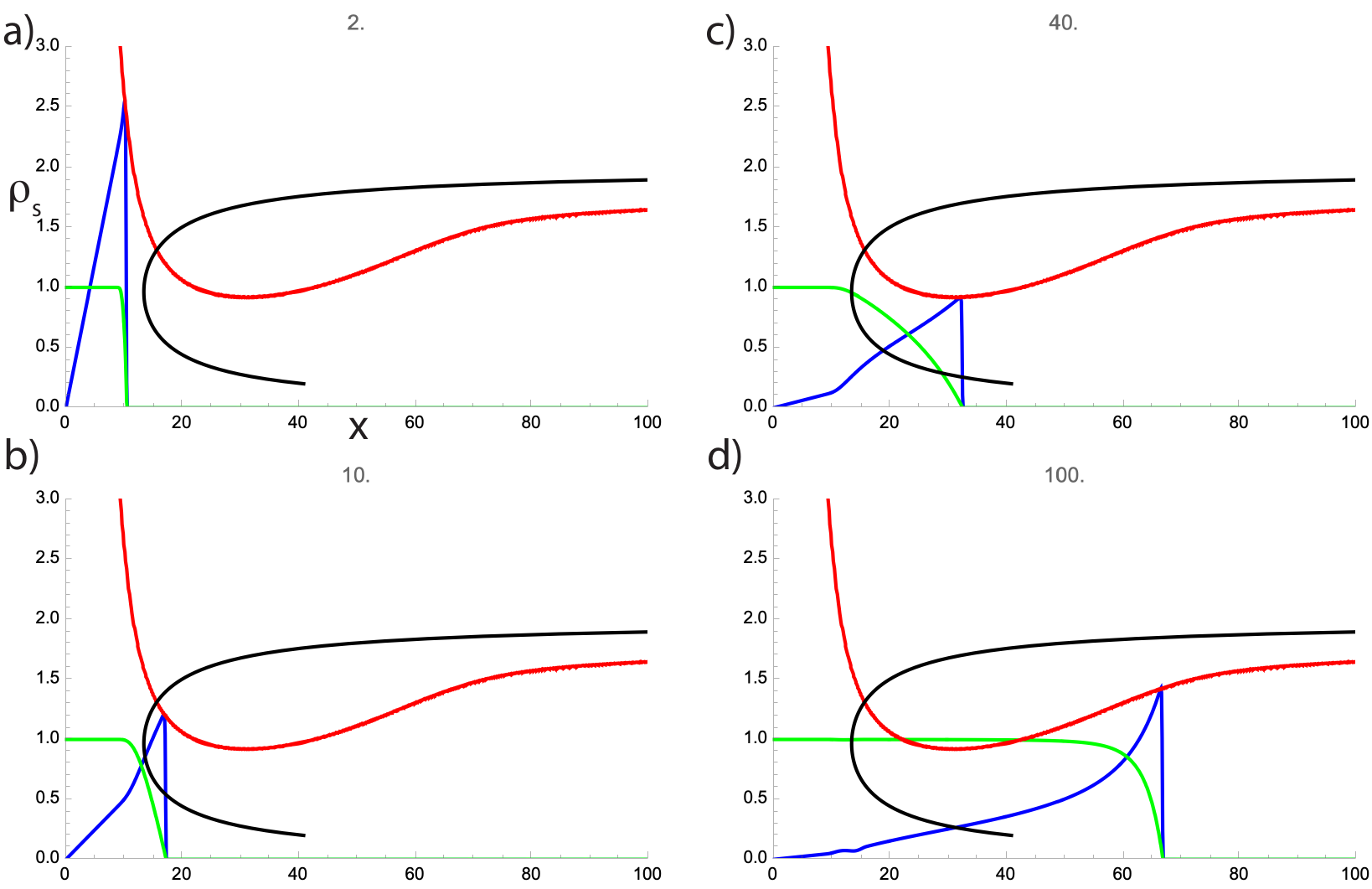}
	\caption{The fields $\rho(x)$ (blue), $\lambda(x)$ (green), $\rho_s(x)$ and the steady response (black) for $x_s = 4$ at a) $t=2$, b) $t=10$, c) $t=40$ and d) $t=100$; see also animations as Supplemental Material.}
	\label{fig:x4}
\end{figure} 
The results obtained further substantiate the He \& Clavin criterion for direct initiation \cite{he_clavin_1994}, whereby the critical energy for initiation can be estimated by determining which inert blast decay curve intercepts the critical turning point of the steady-state response.  Improvements for a better estimate were suggested by Short \& Bdzil \cite{short_bdzil_2003} by using Korobeinikov's blast wave solution perturbed to account for energy release \cite{korobeinikov1991}.

While the short answer is that the He \& Clavin criterion appears to be valid in our case, it is worthwhile to dig deeper for a dynamic justification.  The solution to the reactive Burgers problem provides a simpler setting, which nevertheless incorporates the main physics.  Figures \ref{fig:xts1},\ref{fig:xts2} and \ref{fig:xts4} provide the space-time diagrams of the evolution of three trajectories started at $x_s$ of 1, 2 and 4, bracketing the critical regime.  On the space time diagrams of the evolution of $\rho$ on the colour density plot, we have overlaid the trajectories of the characteristics (red lines) and contours of the reaction progress variable $\lambda$.  These space time diagrams are augmented by the individual profiles of $\rho(x,t)$ and $\lambda(x,t)$ of Figures \ref{fig:x1}, \ref{fig:x2} and \ref{fig:x4} and their animations available as supplemental material. 

The main difference between the subcritical cases ($x_s=1, 2$) and the supercritical  case ($x_s=4$) is the inversion of the density gradient while passing on either side of the steady turning point.  Recall that the density also serves as the speed of forward facing characteristics.  This can be understood as follows.  For the supercritical regime, the blasts starts out with a linear positive density gradient imposed by the self-similar solution.  This gradient decays due to the growth of the blast wave (while remaining linear in the inert solution).  Nevertheless, the reactivity tends to steepen this gradient. As the blast decays and the reactivity decreases, the delay of forward facing characteristics to reach the front increases as the separation between the shock and the bulk of the reaction zone increases and the characteristics' speed decrease. This flattens the gradient.  The subcritical case corresponds to the more rapid decay of the lead shock due to geometrical blast decay than the the sustain of the front by pressure waves having amplified while traversing the reaction zone.  The gradient in this case gets inverted.  The situation is thus the competition between pressure waves amplifying through the reaction zone reaching the shock and the shock decay from geometrical and recession from the center of symmetry. For the super-critical case, the gradient never gets inverted; it does not need to as it re-initiates prior to the inversion of the gradient.   

The gradient of the characteristic speed behind the shock is key here, as it informs on the residence of the pressure waves in the reaction zone, hence their amplification by virtue of \eqref{eq:char1} - see also \cite{radulescu2011nonlinear, tang2013dynamics} for discussion.  When the gradient passes through zero, the characteristics become in phase with the reactivity and build up pressure behind the lead shock.  This picture of shock dynamics being directly controlled by the geometrical effects, reactivity and the gradient of characteristic speeds behind the front can be stated exactly by projecting the reactive Burger's equation \eqref{eq:ficketspherical} along the trajectory of the shock $x_s(t)$.  What amounts to using the so-called shock change equations \cite{radulescushock2020}, the rate of change of the density behind the shock is written as 
\begin{equation}
\frac{\mathrm{d} \rho_s}{\mathrm{d}t}=\left( \frac{\partial \rho}{\partial t}\right)_s +D  \left(\frac{\partial \rho}{\partial x}\right)_s
\end{equation}
The subscript $s$ is to remind us that these are derivatives evaluated immediately behind the shock.  Since the shock jump is $\rho_s=2D$, we can immediately write:
\begin{equation}
\left( \frac{\partial \rho}{\partial t}\right)_s=2 \dot{D}-D  \left(\frac{\partial \rho}{\partial x}\right)_s
\end{equation}
Likewise, the rate of change of the reaction progress variable following the shock motion can be written as:
\begin{equation}
\frac{\mathrm{d} \lambda_s}{\mathrm{d}t}=\left( \frac{\partial \lambda }{\partial t}\right)_s +D  \left(\frac{\partial \lambda}{\partial x}\right)_s
\end{equation}
At the shock, $\lambda$ is constant, so $\frac{\mathrm{d} \lambda_s}{\mathrm{d}t}=0$. We also have the reaction rate controlling the heat release via the time partial derivative via \eqref{eq:rate}.  We can hence write:
\begin{equation}
\left( \frac{\partial \lambda }{\partial x}\right)_s =  -\frac{1}{D}  \left(\frac{\partial \lambda}{\partial t}\right)_s = -\frac{r}{D}
\end{equation}
Using these projections, and the shock jump condition, we can re-write \eqref{eq:ficketspherical} as the sought shock change equation:
\begin{equation}
\frac{\dot{D}}{D}=\frac{1}{4}\frac{Q}{D^2}r-\frac{j}{x}D-\frac{1}{2}\left(\frac{\partial \rho}{\partial x}\right)_s
\end{equation}
This provides a statement of the dynamics of the lead shock, given by the influence of the heat release (first term on the RHS), the geometrical decay (second term on the RHS) and the gradient of characteristic speeds behind the front (third term on the RHS).  It is not yet an evolution equation for the front, since the gradient term needs closure, as it depends on reactivity and the time-history of the solution interior.  

Noting that $\frac{\dot{D}x_s}{D^2}=\frac{\mathrm{d} \ln D}{\mathrm{d} \ln x}$, we can re-write this shock change equation as:
\begin{equation}
\theta\equiv\frac{\mathrm{d} \ln D}{\mathrm{d} \ln x}=\frac{1}{4}\frac{Q}{D^2}\frac{x}{D} r-j-\frac{1}{2} \frac{x}{D} \left(\frac{\partial \rho}{\partial x}\right)_s
\end{equation}
In blast wave theory, $\theta$ is called the blast decay coefficient, and serves to describe the local quasi-self-similar dynamics.  On the basis of this shock change equation, the qualitative description provided above can be put on stronger footing.  At early times, when the energy release term is negligible, the two last terms control the dynamics.  Approximating the derivative in the last term by its linear profile provides the self-similar decay.   Subsequently, the decay of the blast wave is controlled by the change in importance of the 1st and 3rd terms.  The gradient requiring clusure requires the knowledge of the gradients in the characteristic speeds behind the front. 

Without solving the problem, it is sufficient to seek the locus where the characteristic speed gradient vanishes and the zero in blast decay coefficient.  This will be the locus where the reactivity has a strong positive feedback on the gasdynamic amplification and amplification begins.  In the French expression frequently used by my colleague Ashwin Chinnayya to decribe such dynamics: \textit{\c{C}a Pousse!}; which poorly translates to "it starts to grow", or "it starts to self-propel". For a lack of a better nomination, and since Frenchmen like acronyms, this locus will be referred to as the \textit{\c{C}aP} locus.      

Adopting the rate given by \eqref{eq:rate}, the \textit{\c{C}aP} locus is given by 
\begin{equation}
D_{cap}=\left(\frac{4 j}{x Q}\right)^{\frac{1}{n-3}} \label{eq:cap}
\end{equation}

This locus is shown in Fig.\ \ref{fig:logdecays}.  It captures very well the locus where the subcritically initiated detonations start growing again.  This locus separates the slow dynamics attracted by the bottom branch of the steady-state response curve and the top branch of fast dynamics.  

Interestingly, the expression \eqref{eq:cap} also reflects the important role of the sensitivity of the reaction rate to shock strength via the exponent $n$. For $n$ greater than 3, the \textit{\c{C}aP} locus flattens out with increasing $n$.  This signifies that the dynamics of the re-initiation in the subcritical regime become increasingly slower, as the threshold to re-initiate involves ever increasing travel distances to reach it.  This again is in good accord with existing phenomenology.  For $n$ less than 3, the slope of the curve changes sign and the criticality in direct initiation disappears, in good accord with phenomenology and the absence of a turning point in the steady state response.  

The existence of the \textit{\c{C}aP} locus marking the critical re-initiation locus prompts us to formulate a criterion \textit{à la} He \& Clavin and Kasimov \& Stewart to estimate the critical blast strength conducive to rapid re-initiation.  Since the solution requires connecting the trajectories of the inert blast wave with the re-initiation \textit{\c{C}aP} locus, by inspection of Fig.\ \ref{fig:logdecays}, this occurs in the vicinity of the CJ solution. Formally matching them at the CJ solution $D=D_{CJ}=\sqrt{Q}$, we obtain an estimate for the critical source strength:
\begin{equation}
m_{3,crit}=2\pi 8^3Q^{\frac{4-3n}{2}}
\end{equation}

The dynamics modeled by our \textit{\c{C}aP} criterion is close in spirit to the model put forward by Kasimov \& Stewart \cite{kasimov_stewart_2005} and Vidal \cite{vidal2009critical} for critically initiated detonation waves modelled by weak curvature and non-steadiness.  In their model, which they do close approximately, they ascribe importance to the dynamics of a limiting characteristic isolating the dynamics of the front from the back. The description requires the notion of this limiting characteristic to persist in the dynamics.  For the subcritical case of Fig.\ \ref{fig:xts1}, all characteristics feed into the lead shock, albeit after a very long time. For the re-initiating cases of Fig.\ \ref{fig:xts2} and Fig.\ \ref{fig:xts4}, there is a characteristic that never reaches the front present in the reaction zone.  It would be of interest to re-formulate the Kasimov and Stewart model for the simplified Fickett problem presented here in order to assess its accuracy in capturing its dynamics.  Particularly, the model predicts a separatrix.  It is not clear at present how this separatrix connects to our \textit{\c{C}aP} locus obtained on the basis of the shock change equation applied at the shock.  

\section{Conclusion}
In closing, the results obtained using our toy model illustrate well how the reactivity and its sensitivity to the shock controls the speed of steady curved waves and their initiation dynamics.   The existence of a limiting characteristic in the steady problem permits to establish the fraction of energy release "lost" after the sonic surface.  This fraction increases with increasing loss rate. Indeed, it is given by equating the loss rate to the energy release rate at the sonic surface.  The second contribution to the speed deficit and turning point comes from the lengthening of the reaction zone structure in the presence of losses.   In the transient case of direct initiation of a detonation from a strong self-similar blast wave, we have shown how the reactivity competition with geometrical effects control the re-initiation locus, which rapidly extends to very large distances once the detonation decays in the subcritical regime. The importance of a favorable characteristic speed gradient for re-initiation was emphasized.  It would be of great interest to verify whether the present non-steady description applies to the transient case of the Euler equations.      

\section*{Acknowledgments} 
This work was supported by AFOSR grant FA9550-23-1-0214, with Dr.\ Chiping Li as program monitor.  The author also acknowledges financial support from NSERC Discovery Grant "Predictability of detonation wave dynamics in gases: experiment and model development".

\section*{Supplementary material} 
Video animations of the results presented in Figs.\ \ref{fig:x1}, \ref{fig:x2} and \ref{fig:x4} are provided as supplementary material.

------------------------------------------------------------ %

\bibliographystyle{IEEEtran}
\bibliography{references}


\end{document}